\documentclass[multphys,vecphys]{svmult}

\usepackage{makeidx}         
\usepackage{graphicx}        
\usepackage{multicol}        
\usepackage[bottom]{footmisc}

\makeindex             

\begin{document}

\title*{High resolution spectroscopy of HgMn stars: a time of surprises}
\author{
S.\ Hubrig\inst{1},
C.R.\ Cowley\inst{2},
F.\ Gonz\'alez\inst{3},
F.\ Castelli\inst{4}}
\authorrunning{Hubrig et~al.}
\institute{
European Southern Observatory, Casilla 19001, Santiago, Chile 
\texttt{shubrig@eso.org},
\and University of Michigan, Ann Arbor, MI 48109-1042, USA
\texttt{cowley@umich.edu}
\and Complejo Astron\'omico El Leoncito, Casilla 467, 5400 San Juan, Argentina 
\texttt{fgonzalez@casleo.gov.ar},
\and Osservatorio Astronomico di Trieste, Trieste, Italy
\texttt{castelli@ts.astro.it}}
%
%
\maketitle

The origin of the abundance anomalies observed in late B-type stars with HgMn 
peculiarity is still poorly understood. 
Observationally, these stars are characterized
by low rotational velocities and weak or non-detectable magnetic fields.
The most distinctive features of their atmospheres are an extreme
overabundance of Hg (up to 6 dex) and/or Mn (up to 3 dex) and a deficiency of He.
Anomalous isotopic abundances have been reported in the past for the elements Hg, Pt and Tl. 
Observational evidence for large isotopic shifts in the infrared triplet of Ca~II
was presented in the last two years (Castelli \& Hubrig 2004 \cite{CastelliHubrig04}, 
Cowley \& Hubrig 2005 \cite{CowleyHubrig05}, Cowley et 
al., these proceedings).
Shifts of up to +0.2\,\AA{} were found in a number of HgMn and magnetic Ap stars indicating the 
dominant isotope is the terrestrially rare $^{48}Ca$.

As more than 2/3 of the HgMn stars are known
to belong to spectroscopic binaries (Hubrig \& Mathys 1995 \cite{HubrigMathys95}),
the variation of spectral lines observed in any HgMn star is usually explained to be due to
the orbital motion of the companion. 
Here we present the results of a high spectral resolution study of a few spectroscopic
binaries with HgMn primary stars.
We detect for the first time  in the spectra of HgMn stars that for many elements 
the line profiles are variable over the rotation period (Hubrig et al. 2006 \cite{HubrigGonzalez06}).
The strongest profile variations are found for the elements Pt, Hg, Sr, Y, Zr, Mn, Ga, He and Nd.
The slight variability of  He and Y is also confirmed by the study of high resolution spectra
of another HgMn star, $\alpha$\,And. 



\begin{figure}
\centering
\includegraphics*[width=0.46\textwidth,clip=]{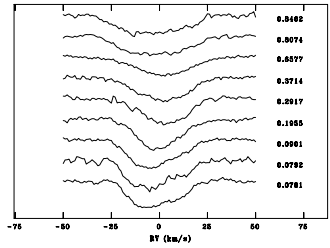}
\caption{
Variations of the line profile of Hg\,II $\lambda$\,3983.9 in the 
spectra of AR\,Aur phased with the rotation period P = 4.13\,days.
}
\label{hubrig2:fig1}       
\end{figure}

\begin{figure}
\centering
\includegraphics*[width=0.45\textwidth,clip=]{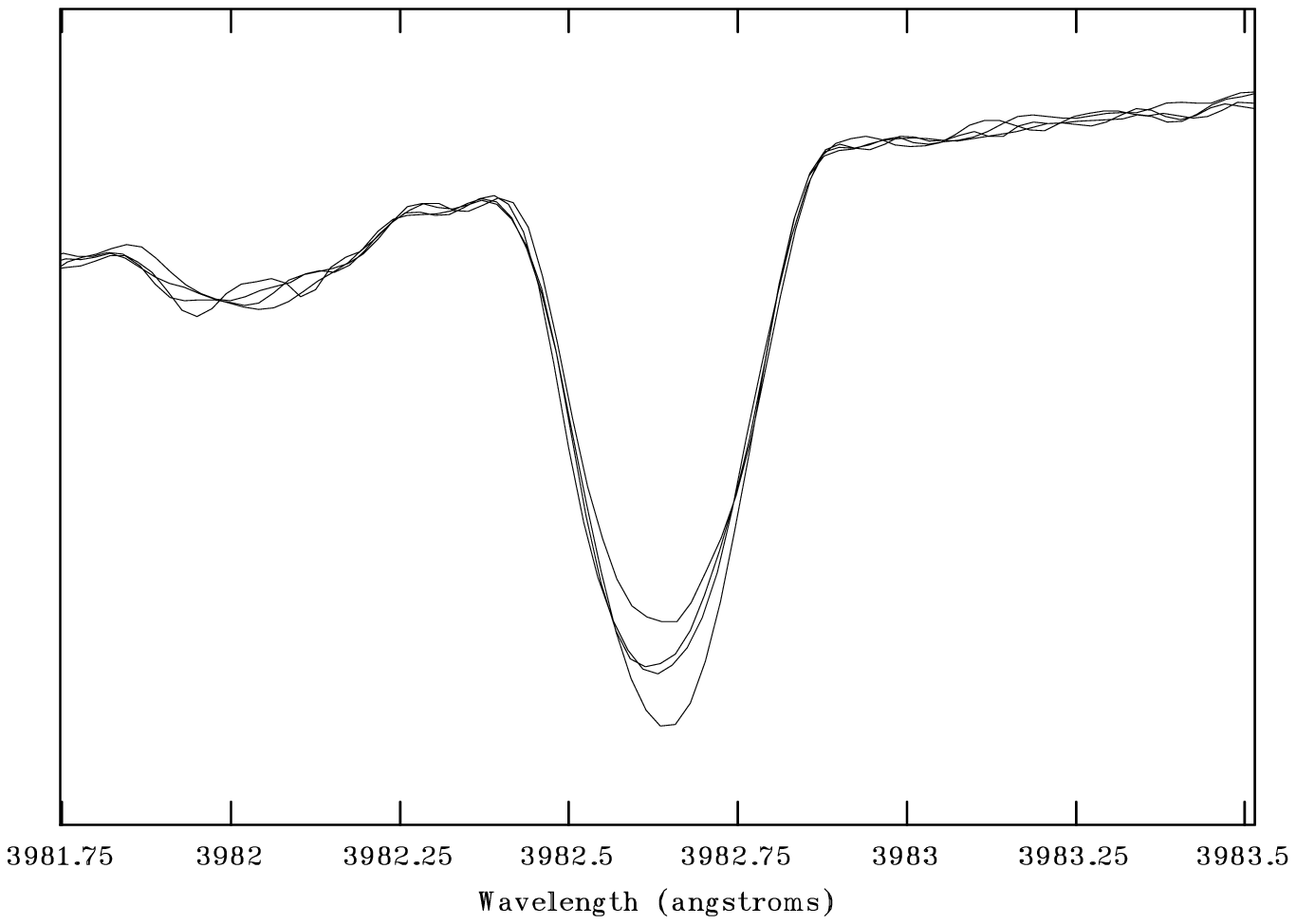}
\includegraphics*[width=0.45\textwidth,clip=]{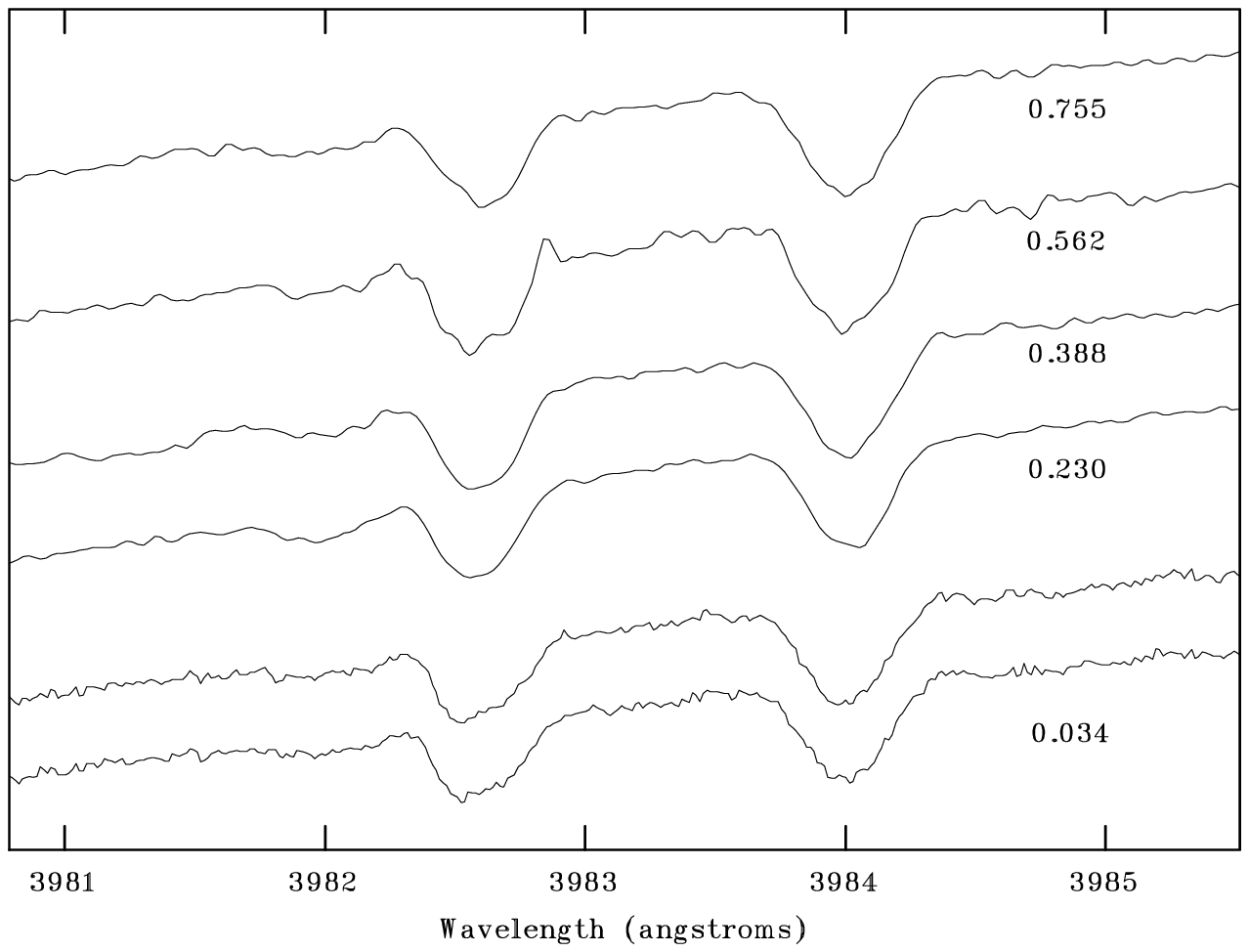}
\caption{
Left: Profile variations of the Y~II $\lambda$\,3982.5\,\AA{} line in the spectra of HD\,11753
at various rotation phases.
Right: Profile variations of the Y~II $\lambda$\,3982.5 and Hg\,II $\lambda$\,3983.9 lines in 
the spectra of HD\,32964 over the rotation period.
}
\label{hubrig2:fig2}       
\end{figure}

In Fig.~\ref{hubrig2:fig1} we show the behavior of the line
profile of Hg\,II $\lambda$\,3983.9 in the spectra of 
AR\,Aur at different rotation phases. Our 
preliminary modelling of abundance distributions of the elements Sr and Y over the stellar surface
suggests that these elements are very likely concentrated in a fractured ring along the 
rotational equator (Hubrig et al. 2006 \cite{HubrigGonzalez06}).
In Fig.~\ref{hubrig2:fig2} we present recent FEROS observations of variable 
line profiles of Y~II $\lambda$\,3982.5 in the 
spectra of the HgMn star HD\,11753 and of Y~II/Hg~II lines in the spectra of the 
HgMn double-lined spectroscopic binary HD\,27376.

The discovery of an inhomogeneous distribution of various elements in the atmospheres of HgMn stars
challenges our understanding of the nature of HgMn stars.
We believe that factors as the presence of a weak tangled magnetic field, tidal distortion, or 
the reflection effect can play a role in the development of anomalies in HgMn stars. 
Although diffusion due to gravitational settling and radiative levitation remain the most popular 
explanation for the HgMn star abundances, the selective accretion of interstellar material 
during the pre-main-sequence phase by HgMn binary systems seems to be a promising possibility 
for explaining some surface anomalies, given the presence of magnetic fields.
Probably, all these 
mechanisms have to be taken into account in future studies of these stars.



\end{document}